# AOC: <u>A</u>nalysis of <u>O</u>rthologous <u>C</u>ollections - an application for the characterization of natural selection in protein-coding sequences

**Authors** Alexander G. Lucaci[1,2], Sergei L Kosakovsky Pond[3]

**Institutions**
[1]Department of Physiology and Biophysics, Weill Cornell Medicine, Cornell University, New York, NY 10021, USA
[2]The HRH Prince Alwaleed Bin Talal Bin Abdulaziz Alsaud Institute for Computational Biomedicine, Weill Cornell Medicine, New York, NY 10021, USA
[3] Institute for Genomics and Evolutionary Medicine, Temple University, Philadelphia, PA, USA

## Abstract

**Motivation**

Modern molecular sequence analysis increasingly relies on automated and robust software tools for interpretation, annotation, and biological insight. The Analysis of Orthologous Collections (AOC) application automates the identification of genomic sites and species/lineages influenced by natural selection in coding sequence analysis. AOC quantifies different types of selection: negative, diversifying or directional positive, or differential selection between groups of branches. We include all steps necessary to go from unaligned homologous sequences to complete results and interactive visualizations that are designed to aid in the useful interpretation and contextualization.

**Results**

We are motivated by a desire to make evolutionary analyses as simple as possible, and to close the disparity in the literature between genes which draw a significant amount of interest and those that are largely overlooked and underexplored. We believe that such underappreciated and understudied genetic datasets can hold rich biological information and offer substantial insights into the diverse patterns and processes of evolution, especially if domain experts are able to perform the analyses themselves.

**Availability and implementation**

A Snakemake [Mölder et al., 2021] application implementation is publicly available on GitHub at https://github.com/aglucaci/AnalysisOfOrthologousCollections and is accompanied by software documentation and a tutorial.

## Introduction

Genomic research is inevitably biased towards certain organisms (humans, model organisms, agriculturally important species, pathogens), and genes (biomedically important, functionally understood) [Stoeger et al, 2018]. For example, GeneRif -- a database of the reference set of articles describing the function of a gene [https://www.ncbi.nlm.nih.gov/gene/generif-stats, last accessed July 6, 2023], is dominated by 5 species: Humans, Mouse, Rat, Arabidopsis, Drosophila corresponding to about 92% of total

# AOC: <u>A</u>nalysis of <u>O</u>rthologous <u>C</u>ollections - an application for the characterization of natural selection in protein-coding sequences

coverage; Humans alone represent 63% of all GeneRifs. This highly skewed coverage of gene-level functional information, concentrated in a largely anthropocentric fashion, fails to benefit from the knowledge gained through the diversity of genetic form and function in the natural world.

The AOC application is designed to be a one stop shop for molecular sequence evaluation using state of the art methods and techniques. The pipeline is fully automated and incorporates recombination detection, a powerful force in shaping gene evolution which can produce spurious results if not taken into account. The application is simple to install and use, requiring few dependencies and few input files or configuration. We differentiate ourselves from other approaches in the field [Picard et. al., 2020] by data preparation steps we take (see Figure 1), and the selection analysis modalites we take advantage of which include lineage-specific and site-level information, and search for pervasive or episodic selective patterns with consideration of positive, negative, directional, biochemical, between-group comparison, and relaxed evolutionary forces. As an example application of AOC, we were able to report on novel sites of adaptive evolution, broad relationships of coevolution, and independently verify previously reported results on the signatures of purifying selection in the mammalian BDNF [Lucaci et. al., 2022] gene, which plays a critical role in brain development.

We are also motivated by the so-called "day science" and "night science" [Yanai and Lercher, 2019] scientific duality. Here, "day science" is the application and evaluation of *a priori* hypotheses which are validated or falsified by the available data. We apply this kind of evaluation because each of the selection analysis methods we use are designed to ask and answer particular biological and statistical questions (we highlight these in the Implementation section). However, we also focus on "night science" where a user can explore the *"unstructured realm of possible hypotheses, of ideas not yet fully fleshed out"* [Yanai and Lercher, 2019] which may not have occurred to the user when they first set out to evaluate their gene of interest. Therefore, AOC is designed as a blend between the two philosophical lines of inquiry, where a user can approach the application with a particular hypothesis in mind, but also allows for data exploration to serve as a guide on a scientific adventure not previously considered. In addition, as the AOC application use grows, the results of particular experiments can become part of a kind of genetic profile, allowing for placement in a repository and subsequent meta-analysis.

**The *AOC* workflow**

*Data Retrieval and Cleaning*

We query the NCBI Gene database via www.ncbi.nlm.nih.gov/gene and retrieve gene orthologs. This can be done on the basis of a single sequence per species, which is recommended if multiple transcripts are available, in order to limit data bias. Depending on study design we may also limit our search to only include species specific taxonomic groups (birds, turtles, lizards, mammals, etc). These queries return full gene transcript (RefSeq transcript) and protein sequence (RefSeq protein) files with tabular data (CSV-format) containing useful metadata (including NCBI accession numbers). Other sources of genomic information can also be used.

# AOC: <u>A</u>nalysis of <u>O</u>rthologous <u>C</u>ollections - an application for the characterization of natural selection in protein-coding sequences

We use protein sequences and full gene transcripts to derive coding sequences (CDS) via a custom script: `scripts/codons.py`. We also recommend using only high quality protein sequences, as "PREDICTED" or "PARTIAL" sequence files may contain errors and are not appropriate for downstream selection analysis. Our application removes low-quality protein sequences from downstream analysis, as they may inflate rates of nonsynonymous change or otherwise bias the analyses.

*Selection analyses*

The AOC application is designed for comprehensive protein-coding molecular sequence analysis. AOC allows for the inclusion of recombination detection, which is a powerful force in shaping gene evolution and critically important to correctly interpreting analytic results which are vulnerable to changing recombinant topologies. We also include an automated method for lineage assignment and annotation which relies on input tabular data (e.g. from NCBI Gene) and NCBI Taxonomy information. Lineage assignment allows for between-group comparisons of selective pressures using selection analysis.

The application accepts two input data files: a protein sequence unaligned `FASTA` file, and a transcript sequence unaligned `FASTA` file for the same gene. Typically, this can be retrieved from public databases such as NCBI Gene (described above). Although this is the recommended route, other methods of data compilation are also acceptable. If protein sequence and transcript sequence files are provided, a custom script `scripts/codons.py` is executed and returns a CDS `FASTA` file. Note that the application is easily modifiable to accept a single CDS input, if such data are available to the user. This script is currently set to assume the standard genetic code, this can be modified for alternate codon tables. This script also removes low-quality sequences (including those where no match is found). The major steps of the AOC pipeline are highlighted below.

> ***Step 1. Codon-aware alignment.*** To generate multiple sequence alignments, we use the Hypothesis testing using Phylogenies (HyPhy) [Pond et. al., 2020] codon-aware multiple sequence alignment procedure available at (https://github.com/veg/hyphy-analyses/tree/master/codon-msa). This procedure is facilitated through the use of MAFFT [Katoh and Standley, 2013] and also has several additional relevant features including frameshift correction and the ability to perform reference-based alignments. We also measure the Tamura-Nei 1993 (TN93) genetic distance of alignments using [https://github.com/veg/tn93]

> ***Step 2. Recombination detection.*** This step is automatically performed using Genetic Algorithm for Recombination Detection (GARD) [Pond et. al., 2006]. A recombination-free set of alignment fragments is placed in the results folder where phylogenetic tree inference and downstream selection analysis are performed. For datasets where recombination is not detected this results in a single file for analysis. In datasets where recombination is detected, we parse out recombinant partitions into multiple files using the software in `notebooks/GARD_parse.ipynb` which corrects for recombinant breakpoints which occur within the border of a codon.

# AOC: <u>A</u>nalysis of <u>O</u>rthologous <u>C</u>ollections - an application for the characterization of natural selection in protein-coding sequences

**Step 3. Phylogenetic tree inference and selection analyses.** For all of the recombination-free `FASTA` files, we perform maximum-likelihood (ML) phylogenetic inference via `IQ-TREE` [Minh et. al., 2020]. Next, the recombination-free alignment and unrooted phylogenetic tree is evaluated through a suite of molecular evolutionary methods designed to ask and answer specific biological and statistical questions including (Spielman et. al., 2019):

- FEL: locates codon sites with evidence of pervasive positive diversifying or negative selection, addresses the question: Which site(s) in a gene are subject to pervasive, i.e., consistently across the entire phylogeny, diversifying selection. [Pond and Frost, 2005]
- BUSTED[S]: tests for gene-wide episodic selection, asks whether a given gene has been subject to positive, diversifying selection at any site, at any time. [Wisotsky et. al., 2020].
- MEME: locates codon sites with evidence of episodic positive diversifying selection, addresses a more general question: Which site(s) in a gene are subject to pervasive or episodic, i.e., only on a single lineage or subset of lineages, diversifying selection [Murrell et. al., 2012].
- aBSREL: tests if positive selection has occurred on a proportion of branches, asks whether some proportion of sites is subject to positive selection along specific branches or lineages of a phylogeny [Smith et. al., 2012].
- SLAC: performs substitution mapping, addresses the question: Which site(s) in a gene are subject to pervasive, i.e., consistently across the entire phylogeny, diversifying selection [Pond and Frost, 2005].
- BGM: identifies groups of sites that are apparently co-evolving.[Poon et. al., 2008].
- RELAX: compare gene-wide selection pressure between the query clade and background sequences, is there evidence that the strength of selection has been relaxed (or conversely intensified) on a specified group of lineages relative to a set of reference lineages [Wertheim et. al., 2014].
- CFEL: comparison site-by-site selection pressure between query and background sequences, [Pond et. al., 2020].
- FMM: examines model fit by permitting multiple instantaneous substitutions, [Lucaci et al., 2021].
- FUBAR, addresses the question: Which site(s) in a gene are subject to pervasive, i.e., consistently across the entire phylogeny, diversifying selection [Murrell et. al., 2013].
- BUSTED[+S+MH]: tests for gene-wide episodic selection while accounting for synonymous rate variation and multiple instantaneous substitutions [Lucaci et al, 2023].

**Step 4. Lineage assignment and tree annotation.** For all of the unrooted phylogenetic trees, we perform automated lineage annotation via the NCBI and the python package `ete3` toolkit [Huerta-Cepas et. al., 2016]. Lineages are binned into K (by default, K = 20)

# AOC: <u>A</u>nalysis of <u>O</u>rthologous <u>C</u>ollections - an application for the characterization of natural selection in protein-coding sequences

number of taxonomic groups. Here, the aim is to have a broad representation of taxonomic groups, rather than the species being heavily clustered into a single group. As a reasonable approximation, we aim for <40% of species to be assigned to any one particular taxonomic group. We perform tree labeling via the `hyphy-analyses` script `Label-Trees` method and results in one annotated tree with a designation for all lineages.

- [https://github.com/veg/hyphy-analyses/tree/master/LabelTrees](https://github.com/veg/hyphy-analyses/tree/master/LabelTrees)

**Step 5. Selection analyses on lineages.** The recombination-free fasta file and the set of annotated phylogenetic trees (where labeling was performed in Step 4) is provided for analysis with the RELAX and Contrast-FEL methods.

**Step 6. Visualization, tables, figure legends.** We provide a high level executive summary and multiple-test correction of the selection analyses and on input files where available for information such as sequence divergence. In addition, we generate figures from all selection analyses along with accompanying summary result tables and figure legends which describe the results. Individual results, specifically output JSON files from HyPhy analyses may also be visualized using Hyphy-Vision [[http://vision.hyphy.org](http://vision.hyphy.org)] or interactive ObservableHQ [Perkel 2021] notebooks [[https://observablehq.com/@hyphy](https://observablehq.com/@hyphy)].

# AOC: <u>A</u>nalysis of <u>O</u>rthologous <u>C</u>ollections - an application for the characterization of natural selection in protein-coding sequences

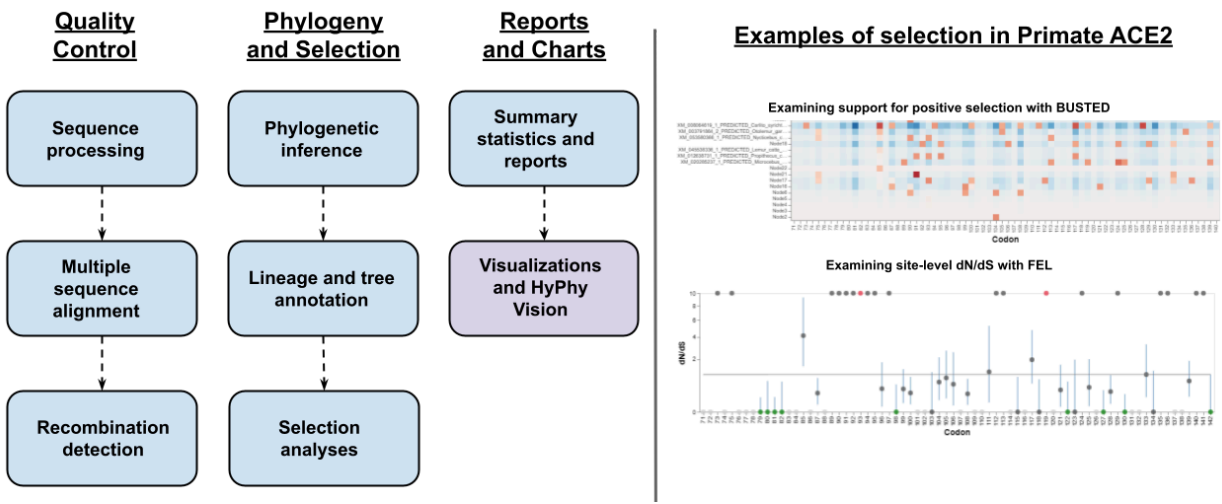

**Figure 1. A flowchart diagram of the AOC workflow and an example using Primate ACE2 data.** The workflow consists of three parts, the first of which does quality control, and converts input transcript and protein files from the NCBI ortholog database into codon-aware alignments and checks for phylogenetic evidence of genetic recombination. The second part performs full maximum-likelihood phylogenetic inference and lineage annotation based on NCBI Taxonomy and runs a full suite of selection detection methods using HyPhy. The last part consists of summarizing results into useful tables and visualizations that can be used for *post-hoc* interpretation and interactions.

**Motivation and example**

We explored the evolutionary history of the primate ACE2 protein. Data was accessed from NCBI via the Ortholog database at https://www.ncbi.nlm.nih.gov/gene/59272/ortholog/?scope=9443&term=ACE2. We downloaded FASTA files from 32 species, with RefSeq Transcripts and RefSeq Proteins (one sequence per species) and metadata in tabular form (CSV). Additional details of our analysis, including all intermediate and HyPhy JSON files are available in our dedicated GitHub repository at https://github.com/aglucaci/AnalysisOfOrthologousCollections/tree/main/results/PrimateACE2.

**Performance evaluation**

For more information on how selection analysis scales along with dataset complexity and size, we refer the reader to HyPhy benchmarking results available in the interactive notebook - https://observablehq.com/@stevenweaver/hyphy-benchmarks-and-profiling.

# AOC: <u>A</u>nalysis of <u>O</u>rthologous <u>C</u>ollections - an application for the characterization of natural selection in protein-coding sequences

**Implementation, flexibility, reproducibility**

The application of modern pipelines for molecular sequence evaluation is of critical importance. These methods have proven to be powerful [Martin et al., 2021, Viana et al., 2022, Tegally et al., 2022, Martin et al., 2022, Bennedorf et al., 2022, Silva et al., 2023, Zehr et al., 2023] to detect the role of natural selection in shaping proteins and offer the ability to further interrogate their results with carefully designed experimental approaches. The combination of computational and experimental biology has the potential to drive significant innovation and discovery in both the basic and translational sciences. With this in mind, AOC is designed to play a role in scientific and medical discovery by providing a simple-to-use software application for molecular sequence analysis especially for insights into unexplored genetic datasets.

## Acknowledgements

We would like to thank members of the HyPhy and Datamonkey teams for their contributions to this project, method development, and the maintenance of state-of-the-art molecular sequence analysis software. This work was supported by a NIH grant (GM151683) to SLKP.

# AOC: <u>A</u>nalysis of <u>O</u>rthologous <u>C</u>ollections - an application for the characterization of natural selection in protein-coding sequences

# AOC: <u>A</u>nalysis of <u>O</u>rthologous <u>C</u>ollections - an application for the characterization of natural selection in protein-coding sequences

**Supplementary data**

Not applicable